\documentclass[12pt,epsf]{article}
\usepackage{graphicx}
\usepackage{epsfig}
\usepackage{amsmath}
\begin{document}
%%%%%%%%%%%%%%%%%%%%%%%%%%%%%%%%%%%%%%%%%%%

\def\a{\alpha}
\def\b{\beta}
\def\c{\varepsilon}
\def\d{\delta}
\def\e{\epsilon}
\def\f{\phi}
\def\g{\gamma}
\def\h{\theta}
\def\k{\kappa}
\def\l{\lambda}
\def\m{\mu}
\def\n{\nu}
\def\p{\psi}
\def\q{\partial}
\def\r{\rho}
\def\s{\sigma}
\def\t{\tau}
\def\u{\upsilon}
\def\v{\varphi}
\def\w{\omega}
\def\x{\xi}
\def\y{\eta}
\def\z{\zeta}
\def\D{\Delta}
\def\G{\Gamma}
\def\H{\Theta}
\def\L{\Lambda}
\def\F{\Phi}
\def\P{\Psi}
\def\S{\Sigma}

\def\o{\over}
\def\beq{\begin{eqnarray}}
\def\eeq{\end{eqnarray}}
\newcommand{\lsim}{\raisebox{0.6mm}{$\, <$} \hspace{-3.0mm}\raisebox{-1.5mm}{\em $\sim \,$}}
\newcommand{\gsim}{\raisebox{0.6mm}{$\, >$} \hspace{-3.0mm}\raisebox{-1.5mm}{\em $\sim \,$}}

\newcommand{\vev}[1]{ \left\langle {#1} \right\rangle }
\newcommand{\bra}[1]{ \langle {#1} | }
\newcommand{\ket}[1]{ | {#1} \rangle }
\newcommand{\EV}{ {\rm eV} }
\newcommand{\GeV}{ {\rm GeV} }
\newcommand{\TEV}{ {\rm TeV} }
\def\diag{\mathop{\rm diag}\nolimits}
\def\Spin{\mathop{\rm Spin}}
\def\SO{\mathop{\rm SO}}
\def\O{\mathop{\rm O}}
\def\SU{\mathop{\rm SU}}
\def\U{\mathop{\rm U}}
\def\Sp{\mathop{\rm Sp}}
\def\SL{\mathop{\rm SL}}
\def\tr{\mathop{\rm tr}}

\def\IJMP{Int.~J.~Mod.~Phys. }
\def\MPL{Mod.~Phys.~Lett. }
\def\NP{Nucl.~Phys. }
\def\PL{Phys.~Lett. }
\def\PR{Phys.~Rev. }
\def\PRL{Phys.~Rev.~Lett. }
\def\PTP{Prog.~Theor.~Phys. }
\def\ZP{Z.~Phys. }

%%%%%%%%%%%%%%%%%%%%%%%%%%%%%%%%%%%%%%%%%%%%%%%%%%%%%%%%%%%%%%%%%%%%

\baselineskip 0.7cm

\begin{titlepage}

\begin{flushright}
UT-09-16
\\
IPMU-09-0067
\end{flushright}

\vskip 1.35cm
\begin{center}
{\large \bf
   A Test for Light Gravitino Scenario at the LHC
}
\vskip 1.2cm
Satoshi Shirai$^{1,2}$ and T. T. Yanagida$^{2,1}$
\vskip 0.4cm

{\it $^1$  Department of Physics, University of Tokyo,\\
     Tokyo 113-0033, Japan\\
$^2$ Institute for the Physics and Mathematics of the Universe, 
University of Tokyo,\\ Chiba 277-8568, Japan}

\vskip 1.5cm

\abstract{
Supersymmetric (SUSY) standard models in which the lightest SUSY particle (LSP) is 
an ultralight gravitino ($m_{3/2}={\cal O}(1)$ eV) are very 
attractive, since they are free from the cosmological gravitino problems.
If the neutralino is the next lightest SUSY particle (NLSP), 
it decays into a photon and the gravitino in collider experiments.
We propose a simple test for the lightness of gravitino at the LHC.
}
\end{center}
\end{titlepage}

\setcounter{page}{2}

\section{Introduction}
Supersymmetric (SUSY) standard models (SSM), 
are the most promising scenario for the beyond-standard model.
Among of them, gauge mediated SUSY-breaking (GMSB) models \cite{Giudice:1998bp} 
with a light gravitino ($m_{3/2}={\cal O}(1)$ eV) are very attractive, 
since they are free not only from the flavor changing neutral current (FCNC) problem 
but also from the cosmological gravitino problems \cite{Viel:2005qj}.
In many GMSB models, the gravitino is the lightest SUSY particle (LSP) and
the Bino-like neutralino or the ``right-handed" stau is most likely the next lightest SUSY particle (NLSP).
In this letter, we investigate the case of the Bino-like neutralino NLSP scenario.
In this case, collider signatures are characterized by 
the simultaneous production of a 
large missing energy, large $P_{\rm T}$ jets and photons.
These signatures are very specific and 
hence they are regarded as a
strong evidence of the beyond-standard model. 
Within the SUSY framework there is almost no doubt that the gravitino is the LSP and the NLSP is the neutralino.
However, it is not clear that such a signal ensures that the SUSY is the case.
For example, some universal extra dimension (UED) models also predict hard photon signals.
However, the particle corresponding to the LSP would be substantially heavy 
($m={\cal O}(100)$ GeV) in UED models.
Therefore, the confirmation for the lightness of the LSP provides a more convincing evidence of the gravitino LSP in the GMSB.
In this letter, we propose a simple test for the light gravitino scenario at the LHC.

\section{Missing Energy and Photon Momentum}
In many GMSB models, colored SUSY particles have larger production cross sections at the LHC, 
even if they are heavier than other colorless SUSY particles.
A heavy SUSY particle decays into a lighter SUSY particle in sequence and
almost all of the SUSY decay chains reach the lightest Bino-like neutralino $\tilde{\chi}^0_1$, which 
mainly decays into a pair of a gravitino and a photon.
(Recall that we assume the NLSP is the neutralino $\tilde{\chi}^0_1$.)
The decay length of the lightest neutralino is given as
\begin{equation}
c\tau\sim 20~{\mu}{\rm m} \left(\frac{m_{3/2}}{1~{\rm eV}} \right)^2\left(\frac{m_{\tilde{\chi}^0_1}}{100~{\rm GeV}}  \right)^{-5},
\end{equation}
which is smaller than the detector size for $m_{3/2}={\cal O}(1)$ eV, and hence the lightest neutralino $\tilde{\chi}^0_1$ decays inside of the detector.
It is expected that two lightest neutralinos are produced in a SUSY event due to the R-parity conservation.
Therefore, two high $P_{\rm T}$ photons and gravitinos are produced in the SUSY events.
We focus on the two photons events.

Consider the decay process $\chi^0_1 \to \tilde{G}_{3/2} + \gamma$.
In the rest frame of the neutralino $\tilde{\chi}^0_1$, regardless of the gravitino mass, 
we have a relation between momenta of 
the gravitino ($\vec{P}_{\tilde{G}_{3/2}}$) and the photon ($\vec{P}_{\gamma}$) as
\begin{equation}
\vec{P}_{\tilde{G}_{3/2}}=-\vec{P}_{\gamma}.
\end{equation}
If the neutralino $\tilde{\chi}^0_1$ has a non-zero momentum, generally
\begin{equation}
\vec{P}_{\tilde{G}_{3/2}}\ne -\vec{P}_{\gamma}
\end{equation}
for each decay.
However, the distributions of the two momenta ($\vec{P}_{\gamma}$ and $\vec{P}_{\tilde{G}_{3/2}}$) should be identical, 
since the gravitino is almost massless;
\begin{equation}
D(\vec{P}_{\tilde{G}_{3/2}})=D(\vec{P}_{\gamma}),
\end{equation}
where $D(X)$ represents a distribution of variable $X$.
On the other hand, in the case of a heavy LSP (here, LSP means the particle corresponding to the gravitino $G_{3/2}$),
\begin{equation}
D(\vec{P}_{\rm LSP})\ne D(\vec{P}_{\gamma}),
\end{equation}
if the neutralinos have  non-zero momenta.

Therefore, the check of the coincidence of the  distributions
$D(\vec{P}_{\tilde{G}_{3/2}})$ and $D(\vec{P}_{\gamma})$
provides a good test for the smallness of the gravitino mass.
Though the momentum of the individual gravitino cannot be measured directly,
we expect the sum of transverse momenta of two gravitinos are determined by the missing energy $E_{\rm miss,T}$ as,
\begin{equation}
\sum_{i=1,2}P_{\tilde{G}_{3/2}\rm T}^{i} \simeq E_{\rm miss,T}.
\end{equation}
If the gravitino LSP is almost massless, the distribution of the sum of transverse momenta of two photons, 
$|\vec{P}_{ \gamma 1,{\rm T}}+\vec{P}_{\gamma 2,{\rm T}}|$,
 should be identical to that of the missing energy $E_{\rm miss,T}$;
\begin{equation}
D(E_{\rm  miss,T})\simeq D(|\vec{P}_{ \gamma 1,{\rm T}}+\vec{P}_{\gamma 2,{\rm T}}|). \label{eq:dis}
\end{equation}
In the above discussion, we have neglected the missing energy from neutrinos and 
backgrounds related to 
the detector resolution, and hence it is not clear 
whether Eq.~(\ref{eq:dis}) holds at the detector level.
In the following section, we show that the naive relation Eq.~(\ref{eq:dis}) is indeed valid, even if such effects are included.

\subsection*{An Example}

Let us see the shape of $D(E_{\rm miss,T})$  and $ D(|\vec{P}_{ \gamma 1,{\rm T}}+\vec{P}_{\gamma 2,{\rm T}}|)$ 
by adopting an example.
We have done Monte-Carlo simulation to calculate the above two distributions. 
We have used programs Herwig 6.5~\cite{HERWIG6510} and AcerDET-1.0~\cite{RichterWas:2002ch}.
As an example, we adopt the minimal gauge mediation model with
\begin{equation}
\Lambda = 100~{\rm TeV},~M = 200~{\rm TeV},~N_5 = 1,~ \tan\beta = 15,~{\rm sgn}(\mu)=+1~{\rm and}~C_{\rm grav}=1, \label{eq:para}
\end{equation}
where $\Lambda$ is $F/M$, $M$ the messenger mass, $N_5$ the number of messengers, 
$\tan\beta$ ratio of the two higgs vacuum expectation values,
${\rm sgn}(\mu)$ sign of the higgsino mass
and $C_{\rm grav}$ scale factor of the gravitino mass.
Here $F$ is the SUSY breaking F-term.

In the above set up, the gaugino masses $M_a$ are generated 
from loop diagrams of the messengers and are given by, at the one-loop level, 
\begin{equation}
M_{a} \simeq \frac{N_5\alpha_a}{4\pi}\Lambda ~ ~(a=1,2,3),\label{eq:gaugino_mass}
\end{equation}
where $\alpha_a$ is the SM gauge couplings and $\alpha_1=5 \alpha _{\rm EM}/(3 \cos^2\theta_{W})$.
Scalar masses $m_{\phi_i}$, at the two loop level, are given by
\begin{equation}
m^2_{\phi_i}\simeq 2N_5\Lambda ^2 \sum_a \left(\frac{\alpha_a}{4\pi}\right)^2 C_a (i), \label{eq:scalar_mass}
\end{equation}
where $C_a(i)$ are Casimir invariants for the particle $\phi_i$ 
($C_1(i) = 3Y_i^2/5$).
Finally, the gravitino mass is given by
\begin{eqnarray}
m_{3/2} = C_{\rm grav}\frac{F}{\sqrt{3}M_P},
\end{eqnarray}
where $M_P =2.44 \times 10^{18}$~GeV is the reduced Planck mass.

To generate the mass spectrum and decay table, we have used the program ISAJET~7.78~\cite{ISAJET}.
In the above model in Eq.~(\ref{eq:para}), the gravitino mass is 4.8 eV and the neutralino the NLSP of mass 139 GeV.
To eliminate the standard-model (SM) background, the following cuts are imposed;
\begin{itemize}
\item At least four jets with $P_{\rm T1,2}>100$ GeV and $P_{\rm T3,4}>50$ GeV.
\item $E_{\rm miss,T}>100$ GeV.
\item At least two photons with $P_{\rm T}>20$ GeV and $|{\vec{P}_{\gamma 1,{\rm T}}}+{\vec{P}_{\gamma 2,{\rm T}}}|>100$ GeV.
\item No isolated lepton.
\end{itemize}
After those cuts, we find that almost all SM background can be neglected.

\begin{figure}[htbp]
\begin{tabular}{lr}
\begin{minipage}{0.5\hsize}
\begin{center}
\epsfig{file=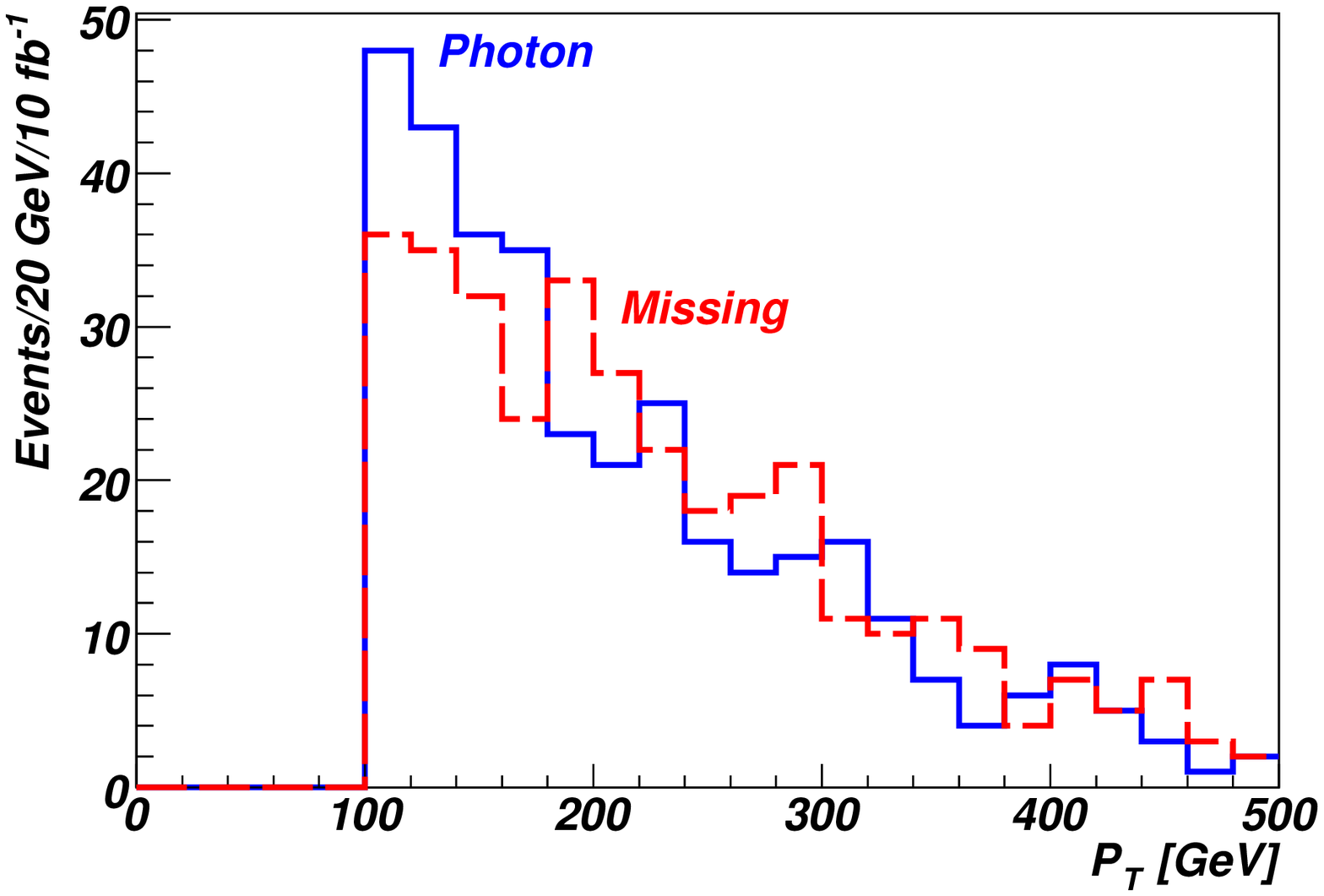 ,scale=.45,clip}
(a)
\end{center}
\end{minipage}&
\begin{minipage}{0.5\hsize}
\begin{center}
\epsfig{file=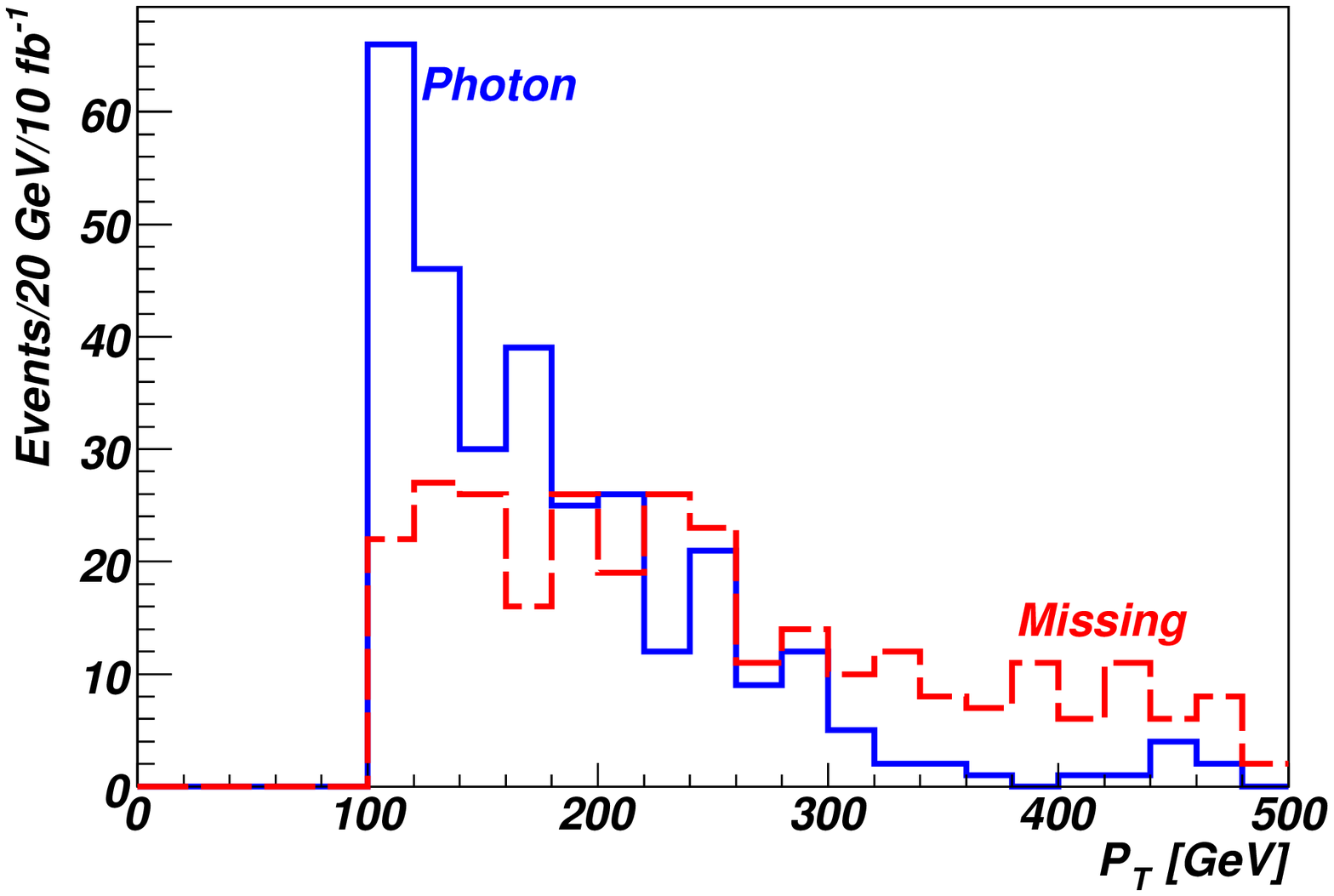 ,scale=.45,clip}
(b)
\end{center}
\end{minipage}
\end{tabular}
\caption{Distributions of the missing energy and the sum of transverse momenta of two photons.
The blue and solid line represents photons' distribution and the red and dashed line the missing energy.
(a): $m_{3/2}=0$ GeV, (b): $m_{3/2}=50$ GeV.
}
\label{fig:eg}
\end{figure}

In Figs.~\ref{fig:eg}, the distributions of transverse missing energy $E_{\rm miss,T}$ and 
the sum of transverse momenta of photons 
$|\vec{P}_{ \gamma 1,{\rm T}}+\vec{P}_{\gamma 2,{\rm T}}|$ are
 shown for an integrated luminosity 10 fb$^{-1}$.
In Fig.~\ref{fig:eg}-(a), the gravitino mass is 4.8 eV.
In Fig.~\ref{fig:eg}-(b), we set the gravitino mass 50 GeV to represent a heavy LSP.
In the latter case, we assume the $\tilde{\chi}^0_1$ decay width into the gravitino is the same as 
in the case that $m_{3/2}=4.8$ eV, 
if it is kinematically allowed.
From Figs.~\ref{fig:eg}-(a) and (b), we see that, in the case of a heavy LSP, the two distributions 
$D(E_{\rm miss,T})$ and $D(|\vec{P}_{ \gamma 1,{\rm T}}+\vec{P}_{\gamma 2,{\rm T}}|)$
have much different forms for each other, 
compared to the massless LSP case.
In the case of the heavy LSP ($m_{\rm LSP}\simeq 50$ GeV), the missing energy tends to be larger than the photon energy.

\section{Scan of Parameter Space}
Here we examine the closeness of the distributions of missing energy and the sum of $\vec{P}_{\gamma}$'s in Eq.~(\ref{eq:dis}) by scanning
for various LSP (``gravitino'') (precisely, LSP is an unknown lightest particle whose interactions 
are assumed to be the same as those of a gravitino of mass $m_{3/2}\approx 1$ eV) mass and various MSSM sparticle spectrums.

We uniformly chose relevant low energy parameters (in GeV) in the following;
\begin{equation}
\begin{matrix}
m_{\tilde g}&\in&[500,1000],~M_{1}&\in&[100,1000],~ M_{2}&\in&[100,1000], \\
m_{{\tilde Q}_1}&\in&[500,1000],~ m_{{\tilde d}_r}&\in&[500,1000],~ m_{{\tilde u}_r}&\in&[500,1000], \\
m_{{\tilde Q}_3}&\in&[500,1000],~ m_{{\tilde b}_r}&\in&[500,1000],~ m_{{\tilde t}_r}&\in&[500,1000], \\
A_{t}&\in&[-500,500],~ A_{b}&\in&[-500,500],~ ~ A_{\tau}&=&0, \\
m_{{\tilde L}_1}&\in&[500,1000],~ m_{{\tilde e}_r}&\in&[500,1000],&&  \\
m_{{\tilde L}_3}&\in&[100,1000],~ m_{{\tilde \tau}_r}&\in&[100,1000],&&  \\
m_{A}&\in&[500,1000],~ \mu&\in&[200,1000].&& 
\end{matrix}
\end{equation}
Here, the second generation sparticles' parameters are the same as the first generation ones.
For the value of $\tan\beta$, we have varied it from 5 to 50 uniformly.
Among them, we have selected mass spectrums which satisfy the conditions that $M_1 < M_{2}, \mu$ and that the neutralino is the NLSP.
We set the LSP (``gravitino") mass 0, 50, 100 and 200 GeV.
As for the decay into the LSP (``gravitino'') 
all the widths of decay into the ``gravitino" are taken as ones of the cases that $m_{3/2}=1$ eV, if it is kinematically allowed, i.e.,
\begin{equation}
\Gamma(X \to {\rm ``Gravitino"}+\sum_{i}Y_i)=
\Gamma(X \to \tilde{G}_{3/2}|_{m_{3/2}=1~{\rm eV}}+\sum_{i}Y_i)
\theta(m_X - \sum_i m_{Y_i}).
\end{equation}
For each LSP (``gravitino") mass parameter, we have generated mass spectrums by using the program ISAJET and
have generated 10000 SUSY events for each mass spectrum by using the event-generator.
We choose the parameter set in which the number of events that survives our cuts are more than 500.
The number of generated mass spectrums and ones which pass our requirement are shown in Table \ref{tb:num}.
To describe the difference between the two distributions, we simply define the quantity as
\begin{equation}
R\equiv \frac{({\rm Mean~ of~} |\vec{P}_{\gamma 1,{\rm T}}+\vec{P}_{\gamma 2, {\rm T}}|)-100~\GeV}{({\rm Mean ~of~}E_{\rm miss,T}) -100~\GeV}.\label{eq:R}
\end{equation}

\begin{figure}[t]
\begin{center}
\epsfig{file=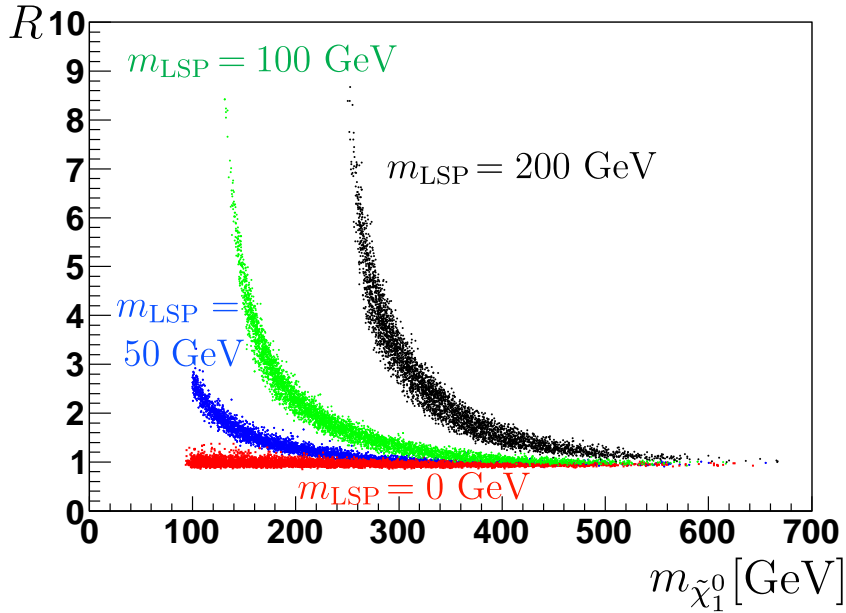 ,scale=.6,clip}
\caption{Scatter plots of $R$ and $m_{\tilde{\chi}^0_1}$ for various values of the LSP mass.
The black dots represent the case that $m_{\rm LSP}=200$ GeV, 
green $m_{\rm LSP}=100$ GeV, blue $m_{\rm LSP}=50$ GeV and red $m_{\rm LSP}=0$ GeV.
The $R$ defined in Eq.~(\ref{eq:R}) is the parameter to represent the difference between missing energy and photon momenta distributions.
}
\label{fig:ratio}
\end{center}
\end{figure}

\begin{table}[h]
\begin{center}
\begin{tabular}{|c|c|c|}

\hline
``Gravitino'' Mass & Generated Mass Spectrum & After the Cuts\\
\hline
0 GeV & 7725 & 7255\\
\hline 
50 GeV & 7618 & 7048\\
\hline
100 GeV & 7159 & 5317\\
\hline
200 GeV & 6523 & 4041\\
\hline

\end{tabular}
\end{center}

\caption{The numbers of generated mass spectrums.
The center column represents spectrums which satisfy the Bino-like NLSP ($M_1 < M_{2}, \m$)
and the right column represents ones in which the number of SUSY events passing the cuts is over 500.
}
\label{tb:num}
\end{table}
In Fig.~\ref{fig:ratio}, we show the scatter plot of $R$ and $m_{\tilde{\chi}^0_1}$.
One see that $R$ tends to be larger for larger LSP mass.
This feature is submerged in the case that the neutralino mass is relatively large compared to the LSP (``gravitino") mass.
If the neutralino $\tilde{\chi}^0_1$ is too heavy,
there is no significant difference between the cases of ``gravitino" of mass 0 and 200 GeV.
However, for the low mass neutralino, the value of $R$ provides a very clear discrimination of the massless gravitino from
heavy LSP (``gravitino")'s.

\section{Conclusion and Discussion}
Two photons + missing energy signature at the LHC is very interesting from  various viewpoints.
First of all, it strongly suggests the light gravitino scenario, in the framework of the SSM. 
In this paper, we show that the two distributions of missing energy and photon momenta is very close to each other, 
if the gravitino is very light.
This provides a simple consistency check for the lightness of the gravitino. 
In our previous work~\cite{Hamaguchi:2008hy}, we showed that the lightest neutralino mass can be measured by using only the missing and photon energies.
However, we assumed that the gravitino is massless in the previous work, 
Thus, we get only the relation between $m_{\tilde{\chi}^0_1}$ and $m_{3/2}$, such as 
$(m_{\tilde{\chi}^0_1}^2 - m_{3/2}^2)/m_{\tilde{\chi}^0_1} $
by using $m_{\rm T2}$ technique \cite{Lester:1999tx}.
In the present paper we propose a simple test for the assumption of the massless gravitino.
In addition, even if the ``gravitino" is massive, we can determine both 
$m_{\tilde{\chi}^0_1}$ and $m_{3/2}$ by using the result of $m_{\rm T2}$ technique and the relation between $R$ and $m_{\tilde{\chi}^0_1}$ 
shown in Fig.~\ref{fig:ratio}.

Although we discuss only the case of neutralino NLSP, the slepton NLSP scenario is also interesting.
In this case, our present analysis would be also applicable, but we need a more careful analysis, since
many leptons are generated through SUSY cascade decays and neutrinos, which have missing energy, 
are substantially produced from the decay of the tau leptons (which originate from NLSP decay.)

In this letter, we have shown that the missing energy tends to be larger,  if the LSP is heavier.
It is expected that this trend would also exist in models other than GMSB ones.
Therefore, the distribution of the missing energy $E_{\rm miss,T}$ would provide important informations on the mass
of the lightest particle (probably a candidate for the dark matter in the universe.)
\section*{Acknowledgements}
We thank Koichi Hamaguchi for helpful discussions and comments.
This work was supported by World Premier International Center Initiative
(WPI Program), MEXT, Japan.
The work of SS is supported in part by JSPS
Research Fellowships for Young Scientists.

\end{document}